\documentclass[pdflatex,sn-mathphys-num]{sn-jnl}

\usepackage{graphicx}
\usepackage{amsmath,amssymb,amsfonts}
\usepackage{xcolor}
\usepackage{textcomp}
\usepackage{booktabs}
\usepackage{lineno}
\usepackage{placeins}
\providecommand{\burl}[1]{\url{#1}}
\hypersetup{hypertexnames=false}

\begin{document}

\title[Early ZrC oxidation by small-cell DFT]{Illuminating early ZrC oxidation using ensemble small-cell DFT with dynamic gas management}

\author*[1]{\fnm{Philip} \sur{Würzner}}\email{pdwurzner@tudelft.nl}

\author[4,1]{\fnm{Boyang} \sur{Chen}}\email{boyangchen@suda.edu.cn}

\author[3]{\fnm{Matthias} \sur{Möller}}\email{m.moller@tudelft.nl}

\author[3]{\fnm{Sebastian} \sur{Feld}}\email{s.feld@tudelft.nl}

\author*[2]{\fnm{Qi-Jun} \sur{Hong}}\email{qhong7@asu.edu}

\author*[1]{\fnm{Yinglu} \sur{Tang}}\email{y.tang-5@tudelft.nl}


\affil*[1]{\orgdiv{Faculty of Aerospace Engineering}, \orgname{Delft University of Technology}, \orgaddress{Delft, The Netherlands}}

\affil[2]{\orgdiv{School for Engineering of Matter, Transport and Energy}, \orgname{Arizona State University}, \orgaddress{Tempe, United States of America}}

\affil[3]{\orgdiv{Faculty of Electrical Engineering, Mathematics and Computer Science}, \orgname{Delft University of Technology}, \orgaddress{Delft, The Netherlands}}

\affil[4]{\orgdiv{School of Optoelectronic Science and Engineering}, \orgname{Soochow University}, \orgaddress{Suzhou, China}}

\abstract{During zirconium carbide oxidation, liberated carbon can remain beneath the oxide or escape as gas, complicating the interpretation of mass measurements. We adapt the small-cell ensemble method using sixteen 140~ps \textit{ab initio} molecular dynamics trajectories at 873--1273~K under a sustained oxygen flux about 300 times that of air, with gas removal. Net mass gain and total carbon transfer into gas or retained chains have apparent activation energies of 0.22 and 0.16~eV. Above 973~K, gas evolution levels off while total carbon transfer increases. The observed fraction of transferred carbon retained in chains rises from 40\% at 973~K to 66\% at 1273~K. Fits after 40~ps suggest declining carbon transfer rates over time, although not in every trajectory. These results indicate that, at this oxygen supply, carbon partitioning during early oxidation depends on temperature, with increasing temperature above 973~K favouring chain formation over gasification.}


\maketitle


\section{Introduction}\label{sec1}

Zirconium carbide ($\mathrm{ZrC}$) is an ultra-high-temperature ceramic (UHTC) considered for thermal protection systems in reusable launch vehicles and nuclear reactors for its ultra-high melting point ($\sim$3800 K), high thermal conductivity, and high-temperature strength \cite{sengupta2019advanced, peterson2023zirconium}. However, poor oxidation resistance severely limits its use in oxygen-containing environments \cite{barnier1988comparative, peterson2023zirconium} and an incomplete understanding of oxidation kinetics limits our ability to ameliorate this drawback. Reported rates and activation energies vary across studies, and although these inconsistencies have recently been unified into successive stages of a single temperature- and pressure-dependent process \cite{Lin2026}, the atomistic mechanisms driving each stage remain unclear, including how carbon retention and gasification affect the protectiveness of the growing oxide.

The oxidation of $\mathrm{ZrC}$ is modeled as proceeding through a zirconium oxycarbide ($\mathrm{ZrC}_x\mathrm{O}_{1-x}$) intermediate \cite{voitovich1974high, Gasparrini2018ZirconiumCarbide, Shimada1990OxidationKinetics, Bartlett1963, RamaRao1994, barnier1988comparative}. Measured limits for the oxygen that the $\mathrm{ZrC}$ lattice can take up on its carbon-sublattice sites without a change in crystal structure, all obtained above 1800~K, range from about 20\% \cite{Rejasse2016, barnier1988comparative} to more than 50\% \cite{Constant1975}. The oxycarbide keeps the structure of the carbide \cite{Sakar1972}, which may explain why this intermediate has proven difficult to observe directly \cite{Gasparrini2018ZirconiumCarbide, RamaRao1994, Belluci2005}. Oxygen substitutes the sublattice carbon, subsequently converting the oxycarbide to zirconium dioxide ($\mathrm{ZrO}_2$), macroscopically described by \cite{Gasparrini2018ZirconiumCarbide}:
\begin{equation}
\mathrm{ZrC} + \frac{1}{2}(1-x)\mathrm{O}_2 \rightarrow \mathrm{ZrC}_x\mathrm{O}_{1-x} + (1-x)\mathrm{C} \label{firstreaction}
\end{equation}
\begin{equation}
\mathrm{ZrC}_x\mathrm{O}_{1-x} + \frac{1}{2}(1+3x)\mathrm{O}_2 \rightarrow \mathrm{ZrO}_2 + x\mathrm{CO}_2, \label{secondreaction}
\end{equation}
where $x$ is the carbon retained in the oxycarbide and $(1-x)$ the carbon liberated.

A few nanometers of surface oxide form on $\mathrm{ZrC}$ in air already at room temperature \cite{Fang2022XPS}. At this temperature, Kato~\textit{et al.}\ identified a disordered, substoichiometric surface oxide following oxygen exposure of $\mathrm{ZrC}(100)$ \cite{Kato2000}. Bulk conversion, however, begins only between ${\sim}483$ and ${\sim}653$~K across powder studies \cite{Cheng2020, Shimada1990OxidationKinetics, Shimada2002, barnier1988comparative}. Above these onsets, oxygen converts the zirconium into $\mathrm{ZrO}_2$ while carbon is retained \cite{Shimada2002, Cheng2020, RamaRao1994, Belluci2005}, and $\mathrm{CO}_2$ evolution follows only 120--210~K higher \cite{Shimada2002}. At the $\mathrm{ZrC}/\mathrm{ZrO}_2$ interface, the retained carbon collects into an amorphous, carbon-rich layer with embedded $\mathrm{ZrO}_2$ nanocrystals \cite{Gasparrini2018ZirconiumCarbide, Shimada1998interface, Shimada1996}. Although monoclinic $\mathrm{ZrO}_2$ is the only stable phase below ${\sim}1450$~K at ambient pressure \cite{Shukla2005}, the nanocrystals at the interface are metastable tetragonal and cubic forms, stabilized by their small size and oxygen deficiency \cite{Shukla2005}, with the retained carbon stabilizing the cubic form \cite{RamaRao1994, Bartlett1963}. They revert to monoclinic at higher temperature, which has been attributed to crystallite growth or to the loss of oxygen vacancies \cite{Gasparrini2018ZirconiumCarbide}. 

Lin~et~al.\ describe up to three successive stages during isothermal oxidation: a protective, diffusion-controlled parabolic stage ($k_1$), a partially protective stage ($k_2$), and reaction-controlled linear oxidation ($k_3$). They associate the loss of protectiveness with $\mathrm{CO}_2$ evolution and scale cracking \cite{Lin2026}. While the scale is intact, the dense inner oxide limits oxygen diffusion and carbon accumulates where the local oxygen supply is depleted \cite{Shimada1996, Shimada2002}. Once cracks pierce the oxide, the carbon-rich interfacial layer becomes the primary barrier \cite{Gasparrini2018ZirconiumCarbide}. Whether oxygen transport through this layer or reaction at the $\mathrm{ZrC}$ surface then sets the rate cannot be decided from the rate law alone, because both produce linear kinetics \cite{Crank1956, Gasparrini2018ZirconiumCarbide}, and the two attributions indeed exist in the literature \cite{Gasparrini2018ZirconiumCarbide, Lin2026}. $^{18}\mathrm{O}$ tracer measurements at 1073~K point to transport, but the extracted diffusivity is confounded \cite{Crank1956} because oxygen simultaneously reacts within the layer to form $\mathrm{ZrO}_2$ nanocrystals and $\mathrm{CO}/\mathrm{CO}_2$ \cite{Gasparrini2018ZirconiumCarbide}.

Temperature and oxygen pressure govern how long each stage persists, or whether it occurs at all. In argon with 2 to 21\,\% $\mathrm{O}_2$ near and below 973~K, Lin~et~al.\ attribute the initial protective stage to carbon retention within a compact oxide scale \cite{Lin2026}. At 1073~K, continuous thermogravimetry of spark-plasma-sintered samples resolves no fully protective stage \cite{Lin2026}, and quenched hot-pressed samples show a single linear regime from 15 minutes onward \cite{Gasparrini2018ZirconiumCarbide}. By 1273--1373~K, the interfacial carbon is itself gasified as $\mathrm{CO}$ or $\mathrm{CO}_2$, leaving a porous, detached monoclinic scale \cite{Gasparrini2018ZirconiumCarbide} that at 1273~K disintegrates to powder \cite{Konnik2024}. Oxygen pressure sets how much carbon is retained, with less remaining as pressure rises \cite{RamaRao1994, Bartlett1963} and interfacial carbon persisting to 1773~K at 20~Pa \cite{Shimada1998interface}. Across these studies, carbon retention is associated with compact scales \cite{Lin2026, Gasparrini2018ZirconiumCarbide} and gasification with cracking and loss of protectiveness \cite{Lin2026}. The stages, however, are characterized only by macroscopic, mass-based kinetics whose activation energies span 70--292~kJ\,mol$^{-1}$ \cite{Kuriakose1964, Bartlett1963, Shimada1990OxidationKinetics, RamaRao1994, barnier1988comparative} and depend on temperature, oxygen pressure, and sample preparation \cite{RamaRao1994, Lin2026}, indicating that such measurements convolve the competing atomic-scale processes that they cannot individually resolve.

Prior density functional theory (DFT) works have characterized 0~K $\mathrm{ZrC}$ oxidation. Oxygen adsorbs dissociatively on $\mathrm{ZrC}$ surfaces \cite{Osei2016} and binds strongly at carbon-adjacent sites across the group~4 carbides \cite{Vines2007}. The C$\leftrightarrow$O surface exchange is exothermic, and carbon leaves as CO, not the $\mathrm{CO}_2$ written in Eq.~\eqref{secondreaction}, only when a second oxygen adatom captures the displaced atom \cite{Rodriguez2005}. Dissolved oxygen occupies carbon vacancies in bulk $\mathrm{ZrC}_x$ \cite{Davey2022}. Finally, $\mathrm{ZrO}_2$ adheres strongly to the carbide through interfacial Zr--O bonds \cite{OseiAgyemang2022_1, OseiAgyemang2022_2}, with the strained oxide relaxing from its cubic to its tetragonal form at the interface \cite{OseiAgyemang2022_1}. Dynamics entered this picture only recently. Kwon~\textit{et al.}\ showed by \textit{ab initio} molecular dynamics (AIMD) that the adsorption, exchange, and substitution steps indeed play out in succession on the pristine (100) surface, with adsorbed oxygen extracting lattice carbon as CO within 130~fs and the vacancies left behind guiding subsequent oxygen substitution into the nascent oxycarbide \cite{Kwon2025}, and Bai~\textit{et al.}\ observed the same rapid dissociative onset on a Zr-containing high-entropy carbide \cite{Bai2025}. Both studies, however, span only 2--5~ps with a single fixed dose of oxygen, so the trajectories end where sustained oxide growth would begin, a restriction Bai~\textit{et al.}\ note explicitly. The competition between carbon retention and gasification thus remains unquantified.

Extending such trajectories within conventional DFT is expensive, and researchers often turn to reactive force fields (ReaxFF) \cite{Newsome2012PI, Newsome2013PII} or machine learning interatomic potentials (MLIPs) \cite{Li2024MLIP} to reach longer times and larger systems. These approaches, however, trade away quantum-mechanical accuracy or generalization to conditions outside their training data \cite{Loew2026}. The timescale problem can instead be addressed within DFT itself. For equilibrium properties, the small-cell coexistence method of Hong and van de Walle showed that an ensemble of short simulations of 100--250-atom cells can replace a single prohibitively large calculation, with the statistics over independent runs providing both the observable and its uncertainty \cite{Hong2013SmallCell, Hong2015Highest}.

We extend the small-cell framework from equilibrium phase coexistence to non-equilibrium surface oxidation. We implement a reactive gas-exchange protocol that sustains oxidation over AIMD timescales by replenishing $\mathrm{O}_2$ and removing gaseous carbon molecules (Section~\ref{SecFluxControl}). We run four independent cells at each of four temperatures spanning 873--1273~K for $\sim$140~ps, over an order of magnitude longer than previous oxidation AIMD, so that trajectory-to-trajectory spread provides the uncertainty on pathway-resolved oxidation kinetics. The trajectories quantify how carbon conversion and its partitioning between retained chains and gas depend on temperature during early oxide growth.

\section{Results}\label{Results}

The small-cell ensemble comprises 16 independent AIMD trajectories across four temperatures (873, 973, 1073, and 1273~K), each run for 140~ps under the reactive gas-exchange protocol of Section~\ref{SecFluxControl}. The cell is initialized with 10 $\mathrm{O}_2$ molecules in a gas phase twice the length of the bulk $\mathrm{ZrC}$, which is itself initialized as a $3\times 2\times 2$ supercell. The time-averaged Hertz--Knudsen impingement fluxes for the four temperature ensembles differed by less than 15~\% (Table~\ref{tab:SimConditions}), providing comparable mean external oxygen supplies. Variation between trajectories is used to assess sampling uncertainty. The calculations are not spin-polarized, which primarily affects the oxygen supply (Section~\ref{SecSpin}). 

\begin{table}[ht]
\centering
\caption{Simulation conditions aggregated over $N=4$ trajectories per temperature, over 140~ps simulations (560~ps per temperature). Pressure and Flux time-averaged means $\pm$ standard deviation.}
\label{tab:SimConditions}
\setlength{\tabcolsep}{10pt}
\begin{tabular}{c c c}
\toprule
\textbf{Temp (K)} & \textbf{Pressure (bar)} & \textbf{Flux $\bar{J}$ ($10^{27}\,\text{m}^{-2}\text{s}^{-1}$)} \\
\midrule
873  & 59.9 $\pm$ 7.4  & 94.4  $\pm$ 11.6 \\
973  & 61.7 $\pm$ 11.2 & 92.1  $\pm$ 16.8 \\
1073 & 73.3 $\pm$ 8.7  & 104.3 $\pm$ 12.4 \\
1273 & 78.2 $\pm$ 3.4  & 102.1 $\pm$ 4.5  \\
\bottomrule
\end{tabular}
\end{table}

\subsection{A partially disordered oxide grows and carbon accumulates}

\begin{figure}[ht]
    \centering
    \includegraphics[width=\textwidth]{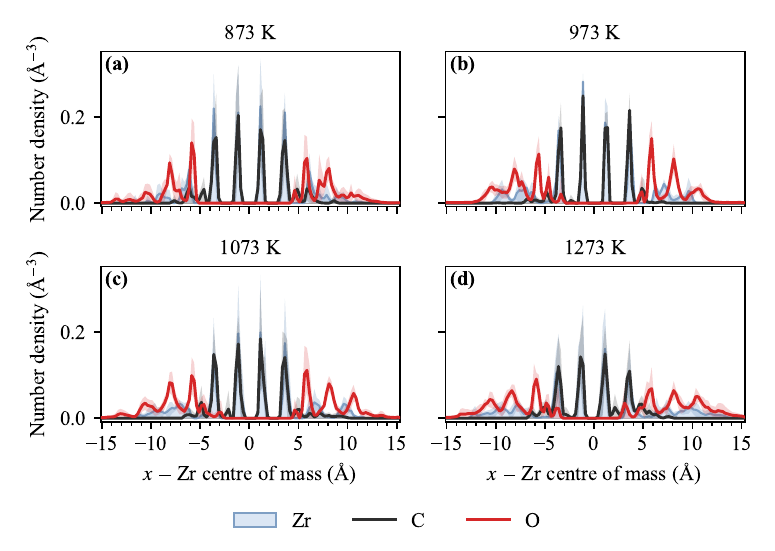}
    \caption{Species number density profiles along the interface normal at
    each temperature, averaged over the final 10~\% (126--140~ps) on a uniform
    50~fs time grid. Grey shaded bands are $\pm 1$~s.d.\ across frames and the
    four independent trajectories.}
    \label{fig:DensityProfiles}
\end{figure}

At all temperatures an oxide forms and carbon accumulates at the interface between the carbide and the oxide (Fig.~\ref{fig:DensityProfiles}). Averaged over the last 10~\% of the simulations (126--140~ps), the oxygen number density shows sharp periodic layering outward from the reaction front. For the 1073 and 1273~K surfaces, the detected oxygen layers approximately follow the carbide plane spacing and phase, while zirconium shows no comparable periodicity across the same region (Supplementary Section~S1). This supports oxygen layering inherited from the carbide, although the density profiles alone do not establish the crystallinity of the oxide. The profiles at earlier times are given in Supplementary Section~S1. Oxygen first forms a single layer on each surface and then spreads both outward and into the outer carbide planes.

\FloatBarrier
\subsection{Minimum carbon retention at 973~K}\label{sec:PhiC}

The conversion of $\mathrm{ZrC}$ into $\mathrm{ZrO}_2$ proceeds through several subreactions. During reaction \eqref{firstreaction} and the conversion of the oxycarbide $\mathrm{ZrC}_x\mathrm{O}_{1-x}$ into $\mathrm{ZrO}_2$,
\begin{equation}
\mathrm{ZrC}_x\mathrm{O}_{1-x} + \frac{1+x}{2}\mathrm{O}_2 \rightarrow \mathrm{ZrO}_2 + x\mathrm{C}, \label{thirdreaction}
\end{equation}
the carbon follows one of two pathways. The first of these pathways is gasification,
\begin{equation}
\mathrm{C} + \frac{y}{2}\mathrm{O}_2 \rightarrow \mathrm{CO}_y, \label{fourthreaction}
\end{equation}
where $y=1$ or 2. The second pathway is retention by carbon coupling into chains,
\begin{equation}
\mathrm{C} + \mathrm{C} \rightarrow 2\mathrm{C}_\mathrm{chain}, \label{fifthreaction}
\end{equation}
which enables a secondary channel for oxidation of carbon:
\begin{equation}
\mathrm{C}_\mathrm{chain} + \frac{y}{2}\mathrm{O}_2 \rightarrow \mathrm{CO}_y. \label{sixthreaction}
\end{equation}
In Eqs.~\ref{fifthreaction} and \ref{sixthreaction}, $\mathrm{C}_\mathrm{chain}$ denotes a single carbon atom bonded in a chain, not a whole chain. Coupling can begin while a carbon atom still occupies its lattice site, so a chain carbon is not necessarily liberated from the $\mathrm{ZrC}$ lattice (Section~\ref{SecDesorptionKinetics}). In our simulations, $\mathrm{CO}$ is the predominant gas product at each temperature (Table~\ref{tab:KineticProducts}). The balance between pathways \eqref{fourthreaction} and \eqref{fifthreaction} may be quantified by the carbon product ratio,
\begin{equation}
    \phi_\mathrm{C} = \frac{n^\mathrm{C}_\mathrm{chain}}{n^\mathrm{C}_\mathrm{chain} + n^\mathrm{C}_{\text{gas}}}, \label{eq:phi_c}
\end{equation}
where $n^\mathrm{C}_\mathrm{chain}$ is the number of carbon atoms currently in chains and $n^\mathrm{C}_{\text{gas}}$ is the cumulative number removed as $\mathrm{CO}$ or $\mathrm{CO}_2$. Similarly, the oxygen retention ratio,
\begin{equation}
    \phi_\mathrm{O} = \frac{n^\mathrm{O}_{\text{scale}}}{n^\mathrm{O}_{\text{scale}} + n^\mathrm{O}_{\text{gas}}}, \label{eq:phi_o}
\end{equation}
where $n^\mathrm{O}_{\text{scale}}$ is the number of oxygen atoms in the oxide scale and $n^\mathrm{O}_{\text{gas}}$ is the cumulative number of oxygen atoms removed in $\mathrm{CO}$ or $\mathrm{CO}_2$.

\begin{table}[ht]
\centering
\caption{Kinetic product details at 140~ps. Product counts are summed over the four trajectories per temperature. Chain length is the number of carbon atoms connected by C--C bonds within the confirmed chain population; the mean and maximum are taken over all such chains pooled across the four trajectories. $\phi_\mathrm{C}$ and $\phi_\mathrm{O}$ are the observed carbon product ratio and oxygen retention ratio (Eqs.~\ref{eq:phi_c}--\ref{eq:phi_o}).}
\label{tab:KineticProducts}
\setlength{\tabcolsep}{5pt}
\small
\begin{tabular}{c cccc cc ccc}
\toprule
 & \multicolumn{4}{c}{\textbf{Reaction products}}
 & \multicolumn{2}{c}{\textbf{$\mathrm{C}_\mathrm{chain}$ Length}}
 & \multicolumn{3}{c}{\textbf{Ratios}} \\
\cmidrule(lr){2-5} \cmidrule(lr){6-7} \cmidrule(lr){8-10}
\textbf{Temp. (K)}
 & $n^{\mathrm{CO}}_{\mathrm{gas}}$
 & $n^{\mathrm{CO}_2}_{\mathrm{gas}}$
 & $n^{\mathrm{O}}_{\mathrm{scale}}$
 & $n^{\mathrm{C}}_{\mathrm{chain}}$
 & \textbf{Avg.}
 & \textbf{Max}
 & $\mathrm{CO}/\mathrm{CO}_2$
 & $\phi_\mathrm{C}$ 
 & $\phi_\mathrm{O}$ \\
\midrule
873  & 21 & 10 & 131 & 35 & 2.69 & 4 & 2.1  & 0.53 & 0.76 \\
973  & 43 & 4  & 163 & 31 & 2.82 & 4 & 10.8 & 0.40 & 0.76 \\
1073 & 36 & 2  & 170 & 48 & 3.00 & 6 & 18.0 & 0.56 & 0.81 \\
1273 & 33 & 3  & 208 & 69 & 3.29 & 8 & 11.0 & 0.66 & 0.84 \\
\bottomrule
\end{tabular}
\end{table}
Table~\ref{tab:KineticProducts} summarizes the product partitioning at 140~ps. At 873~K the carbon pathways are balanced, $\phi_\mathrm{C} = 0.53$ and the formation of $\mathrm{CO}_2$ is higher than at the other temperatures. Among the temperatures studied, only the 973~K ensemble has more gasified than retained carbon, $\phi_\mathrm{C} = 0.40$. At higher temperatures the balance shifts towards retention, with $\phi_\mathrm{C}$ rising to 0.56 at 1073~K and 0.66 at 1273~K, accompanied by longer chains. The single-trajectory data is in accordance with this, with ratios of 0.33 to 0.50 in the four cells at 973~K and 0.57 to 0.76 at 1273~K (Supplementary Section~S2). Oxygen partitioning varies less with temperature. $\phi_\mathrm{O}$ increases from 0.76 to 0.84, so roughly 80\,\% of the reacted oxygen resides in the scale at every temperature, with retention highest where the oxide grows fastest. The ratios calculated in Table~\ref{tab:KineticProducts} are simulation time-dependent. Over 40--140~ps, $\phi_\mathrm{O}(t)$ varies less than $\phi_\mathrm{C}(t)$, which generally shifts toward greater retention (Supplementary Section~S2). Because chain carbon can still gasify (Eq.~\ref{sixthreaction}), the chain count is the carbon in a chain at 140~ps and not the carbon that has entered one. To check for gasification after chain formation, we identified gasified carbons that had previously formed C--C bonds lasting at least 0.2~ps. Of the 152 carbons gasified by 140~ps, 8 had previously entered a chain, while 183 carbons were in chains at 140~ps (Supplementary Table~S10).

\begin{figure}[!ht]
    \centering
    \includegraphics[width=\textwidth]{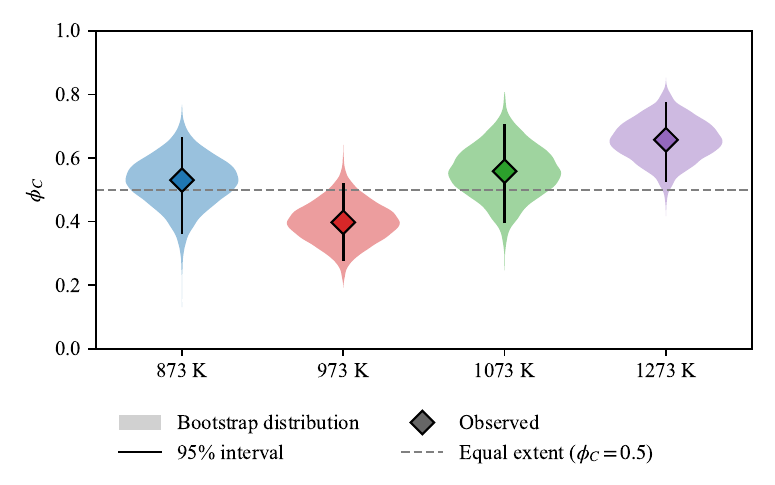}
    \caption{Bootstrap distributions of $\phi_\mathrm{C}$ (Eq.~\ref{eq:phi_c}) at 140~ps. Violins show 10\,000 trajectory resamples per temperature (with replacement, then outcomes resampled within each). Black bars show the 95\,\% confidence interval.}
    \label{fig:PhiCBootstrap}
\end{figure}

To assess uncertainty from the four-trajectory ensembles, we generated 10{,}000 synthetic ensembles per temperature by redrawing the four trajectories with replacement and resampling the carbon products within each draw according to their carbon product distribution (Fig.~\ref{fig:PhiCBootstrap}, Section~\ref{SecPartitioningMethods}). $\phi_\mathrm{C}$ falls below 0.5 in 94~\% of these ensembles at 973~K, versus 34~\%, 23~\%, and 1~\% at 873, 1073, and 1273~K. At 973~K the 95~\% interval of $\phi_\mathrm{C}$ runs from 0.28 to 0.52 and so includes 0.5. Gasification is therefore probably favoured at 973~K, but this is not established. The same is true of the minimum at 973~K (85~\% of resamples) and of the rise from 1073 to 1273~K (83~\%). Only at 1273~K is the outcome established, where chains dominate with an interval of 0.53 to 0.77 (Supplementary Section~S2).

\subsection{Total carbon commitment follows an Arrhenius trend}\label{SecDesorptionKinetics}

At 140~ps roughly half of the lattice carbon remains uncommitted, and the pool of uncommitted carbon depletes as it converts. A rate counted over atoms and elapsed time keeps committed carbons in its denominator after they are gone, understating the rate more where conversion runs further, which would tilt the fitted Arrhenius slope (Supplementary Section~S3). A survival framework removes this bias by assigning each of the 192 lattice carbons per temperature its time at risk, the event time for carbon that commits and the full window for carbon that does not. 

\begin{figure}[ht]
    \centering
    \includegraphics[width=\textwidth]{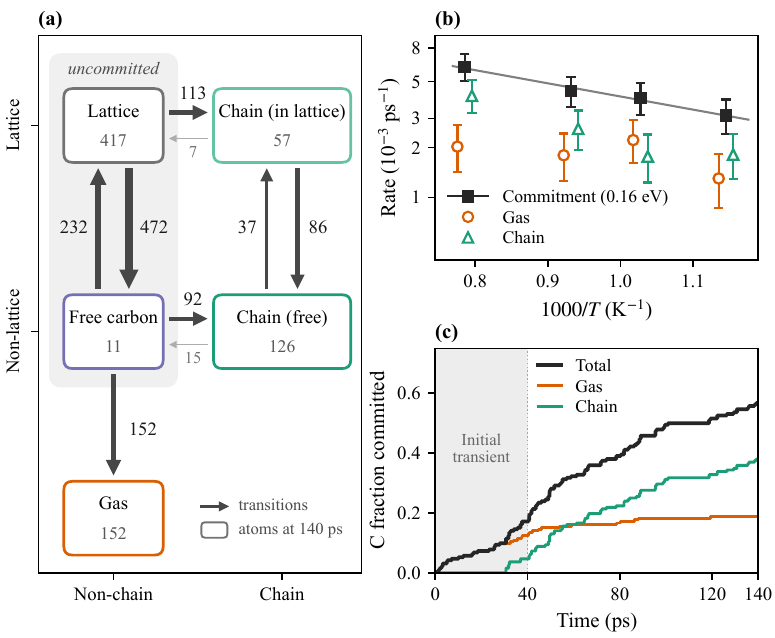}
    \caption{Carbon states and commitment kinetics. (a)~States of the 768 lattice carbon atoms, arranged by zirconium coordination (rows) and carbon bonding (columns). Arrows give the number of transitions summed over the sixteen trajectories, and boxes give the number of atoms in each state at 140~ps. Shading marks the uncommitted states. (b)~Total committed rate $\lambda$ (squares), estimated from all commitment times and times at risk across the full 140~ps window (Section~\ref{SecSurvivalMethods}), with its Arrhenius fit (0.16~eV, solid line), and its gas (circles) and chain (triangles) channels, which sum to it exactly. Error bars are approximate model-based 95~\% intervals of the constant-rate estimates (Eq.~\ref{eq:lambda}). (c)~Cumulative fractions of lattice carbon committed to gas, chains and their total at 1273~K. Shading marks the first 40~ps.}
    \label{fig:Survival}
\end{figure}

Each carbon atom is described by two properties (Fig.~\ref{fig:Survival}a), whether it retains the zirconium coordination of its lattice site and whether it is bonded to another carbon atom. Carbon without a carbon bond is lattice carbon if it has its coordination and free carbon if it has lost it. Bonded carbon is chain carbon, whether or not it still occupies its lattice site (Section~\ref{SecSurvivalMethods}). An atom is liberated when it loses its coordination, and it commits when it forms a C--C bond or desorbs as $\mathrm{CO}$ or $\mathrm{CO}_2$. In our rate analysis, we only count the first commitment of each carbon atom to either gas or chains. This is because the rate is defined per uncommitted atom, so each atom can leave the uncommitted pool only once. Counting the first event also makes the two outcomes mutually exclusive, and it makes the classification of an atom independent of what happens to it afterwards and of the length of the simulation. Later steps, such as a chain carbon that loses its bond and then leaves as gas, appear in Fig.~\ref{fig:Survival}a, which counts every transition and therefore also includes atoms that move back and forth between two states. There are sixteen carbons that form chains and later desorb or lose the bond, so the event counts differ slightly from the product counts in Table~\ref{tab:KineticProducts} (Supplementary Table~S10).

We summarize commitment over the full 140~ps using a constant-rate model, whose maximum-likelihood estimate is
\begin{equation}
\lambda = \frac{n_{\mathrm{ev}}}{\sum_i t_i}, \label{eq:lambda}
\end{equation}
where $n_{\mathrm{ev}}$ counts the commitments to a state and $t_i$ is the observation time of carbon $i$ (time to commit to a gas or chain state or 140~ps if uncommitted). Because the two commitments drain the same pool of uncommitted carbon, the gas and chain rates share one exposure and sum to the total commitment rate. Each estimate carries an approximate model-based 95~\% interval (Supplementary Table~S3, Section~\ref{SecSurvivalMethods}).

The estimated total commitment rate is 3.1, 4.0, 4.4 and $6.1\times10^{-3}$~ps$^{-1}$ at 873, 973, 1073 and 1273~K. A straight-line fit of the logarithm of the rate against inverse temperature gives an apparent activation energy of $0.16 \pm 0.06$~eV ($15 \pm 6$~kJ~mol$^{-1}$, Fig.~\ref{fig:Survival}b), with a 95~\% interval of 0.05 to 0.28~eV. Uncertainties are standard deviations from resampling the trajectories (Supplementary Section~S3). This is an effective value for the simulated conditions.

Commitment is not the same event as liberation. More than half of the transitions into a chain in Fig.~\ref{fig:Survival}a start from lattice carbon (113 of 205), so these C--C bonds form while the carbon is in its lattice site. Liberation is also less sharply defined. Carbon close to the coordination threshold can cross it repeatedly, and the number of such returns to the lattice depends on the parameters used to define states (Supplementary Table~S3c). Of the 558 transitions to a liberated state in Fig.~\ref{fig:Survival}a, 289 do not reverse back into their lattice sites within our simulated window. Of these, 152 end as gas and 126 as chain carbon at 140~ps. The rate of these 289 liberations, estimated in the same way as the commitment rate (Equation~\ref{eq:lambda}), is below the commitment rate at every temperature ($2.6\times10^{-3}$ to $4.7\times10^{-3}$~ps$^{-1}$). Its apparent activation energy of $0.13 \pm 0.06$~eV cannot be distinguished from that of commitment. Ultimately, the commitment rate measures how fast lattice carbon is first converted to gas or chain carbon, and not how fast it leaves the lattice. Because commitment does not depend on the noisier coordination criterion, we use it as the kinetic endpoint.

The gas commitment rate rises from $1.3\times10^{-3}$ to $2.2\times10^{-3}$~ps$^{-1}$ between 873 and 973~K and levels off above 973~K ($1.8\times10^{-3}$ and $2.0\times10^{-3}$~ps$^{-1}$ at 1073 and 1273~K). The chain rate increases from $1.8\times10^{-3}$ to $4.1\times10^{-3}$~ps$^{-1}$ over 873--1273~K. Above 973~K, the additional carbon conversion therefore primarily contributes to retained chains.

We tested whether the rates change over time, since the whole-run estimates average over the development of the oxide and carbon-rich interface. At 1273~K, cumulative commitment grows slowly at first, followed by a faster rise and then more gradual growth (Fig.~\ref{fig:Survival}c). We therefore fitted a Weibull model over 40--140~ps to reduce the influence of the initial transient. Its shape parameter $\rho$ describes how the commitment rate per remaining uncommitted atom changes over time: $\rho < 1$ a declining rate, and $\rho > 1$ an increasing rate. Fits to the combined trajectories give total shapes below one at all four temperatures ($\rho = 0.71$--0.84), suggesting declining commitment rates during this interval. This trend is not shared by every individual trajectory (Supplementary Section~S3). For gas commitment, the fitted shapes range from $\rho = 0.51$ to 0.58 at 973--1273~K, compared with 0.95 at 873~K. Chain commitment gives shapes between 0.79 and 1.03 across all four temperatures. At 973--1273~K, the fits therefore suggest a steeper decline in gas commitment than in chain commitment, consistent with an increasing contribution of chains to later events.

\subsection{Net mass gain and carbon partitioning}\label{SecMacroscopicObservables}

\begin{figure}[ht]
    \centering
    \includegraphics[width=\textwidth]{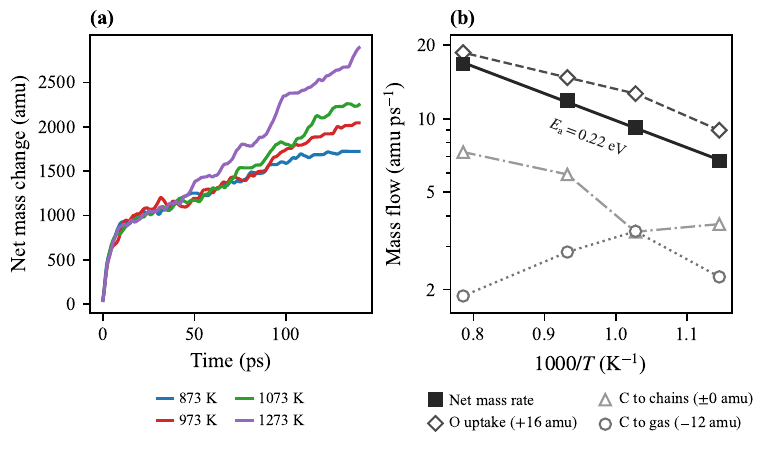}
    \caption{Net mass gain and component flows. (a)~Net mass change of the solid, summed over the four cells per temperature. (b)~The mass flows behind the net rate over 20--140~ps (Section~\ref{SecMacroObsMethods}). The line is the Arrhenius fit to the four net rates ($0.22 \pm 0.04$~eV).}
    \label{fig:MassFlows}
\end{figure}

The net mass of the solid combines oxygen uptake and carbon loss as gas. Each oxygen retained in the scale adds 16.0~amu, while each carbon lost as gas removes 12.0~amu. The resulting signal rises smoothly at every temperature (Fig.~\ref{fig:MassFlows}a). Summed over the four cells, it grows by 6.7, 9.2, 11.9 and 16.8~amu~ps$^{-1}$ at 873, 973, 1073 and 1273~K over 20--140~ps. A straight-line fit of the logarithm of these rates against inverse temperature gives an apparent activation energy of $0.22 \pm 0.04$~eV ($21 \pm 4$~kJ~mol$^{-1}$, Fig.~\ref{fig:MassFlows}b), with a 95~\% interval of 0.15 to 0.29~eV. Four temperatures cannot show that this trend holds outside the range studied. Oxygen uptake increases with temperature and dominates the mass gain, while carbon loss to gas is smaller and peaks at 973~K. More carbon enters chains than gas at three of the four temperatures (Section~\ref{sec:PhiC}), but chain formation does not change the mass of the solid.

\section{Discussion}\label{sec:mechanism}

In our $\mathrm{ZrC}$ oxidation simulations at high pressure, the total carbon commitment rate, averaged over the full 140~ps, increases with temperature. Above 973~K, gas evolution levels off while chain formation increases, shifting the product balance towards retention (Sections~\ref{SecDesorptionKinetics} and~\ref{sec:PhiC}). Faster conversion of lattice carbon therefore does not necessarily lead to faster carbon gas formation. Bartlett similarly described zirconium oxidation as liberating carbon that subsequently burns at its own rate \cite{Bartlett1963}. Oxygen replenishment and gas removal allow us to follow this competition beyond the initial surface reactions accessible in earlier short AIMD studies \cite{Kwon2025, Bai2025}, into a regime where gasification and carbon aggregation proceed together.

The post-40~ps fits suggest declining total commitment rates at all four temperatures, although this trend is not shared by every trajectory (Supplementary Section~S3). Chain commitment typically begins after a delay; at 973--1273~K, the gas rate declines more steeply than the chain rate. As the $\mathrm{Zr}$--$\mathrm{O}$ network grows, it could restrict oxygen access to free carbon or impede its escape, while nearby carbon atoms provide opportunities for C--C bonding. Restricted oxygen access would be consistent with Shimada's attribution of interfacial carbon deposition to low local oxygen activity beneath the oxide \cite{Shimada2002}. Such retention during oxide growth is also observed in ReaxFF simulations of $\mathrm{SiC}$ oxidation, where carbon accumulates alongside the oxide before reacting with oxygen within it \cite{Newsome2012PI}. Preferential consumption of the most accessible lattice carbon could contribute to the declining commitment rate. The fits do not separate this effect from changes in oxygen access, carbon transport or aggregation.

Surface spectroscopy identifies substoichiometric $\mathrm{ZrO}_x$ during initial oxygen exposure, with Kato reporting a disordered surface layer after oxygen dosing at room temperature \cite{Kato2000, Rodriguez2005}. Kato also inferred substrate-related lateral periodicity after annealing the oxygen-dosed surface to 1273~K, which reduced the oxide towards an ordered ZrO-like state \cite{Kato2000}. In our simulations, oxygen layering develops normal to the interface during continued oxidation and, at 1073 and 1273~K, approximately retains the carbide plane spacing (Fig.~\ref{fig:DensityProfiles}). This suggests that the substrate influences the nascent oxide. It does not, however, show that the oxide is crystalline. Beneath the experimental oxide, microscopy and Raman spectroscopy show amorphous carbon with embedded zirconia nanocrystals \cite{Shimada1996, Shimada1998interface, Gasparrini2018ZirconiumCarbide}. The interfacial chains in our simulations may be a precursor to this carbon-rich layer.

Given that oxygen uptake adds mass, gasification removes carbon, and chain formation leaves the mass unchanged, net mass gain alone cannot determine carbon partitioning. Thermogravimetric studies inferred retained carbon from mass gain exceeding that expected for complete conversion to $\mathrm{ZrO}_2$ \cite{Bartlett1963, Shimada2002}. During partial conversion, however, different combinations of oxygen uptake and carbon loss can produce the same net mass. The simulations give a concrete case. The net mass rate follows a single Arrhenius trend between 873 and 1273~K (Fig.~\ref{fig:MassFlows}), while the carbon product ratio $\phi_\mathrm{C}$ varies between 0.40 and 0.66 (Table~\ref{tab:KineticProducts}). A smooth mass curve can therefore conceal a change in the fate of the carbon.

The apparent activation energies are $0.22 \pm 0.04$~eV for net mass gain and $0.16 \pm 0.06$~eV for total carbon commitment (Sections~\ref{SecDesorptionKinetics} and~\ref{SecMacroscopicObservables}), or about 21 and 15~kJ\,mol$^{-1}$. The two cannot be distinguished within the present uncertainty (Supplementary Section~S3). They are effective values for the simulated conditions and do not identify the barrier of a single atomic step. Both lie well below the 70--292~kJ\,mol$^{-1}$ of thermogravimetric studies \cite{Kuriakose1964, Bartlett1963, Shimada1990OxidationKinetics, RamaRao1994, barnier1988comparative}. This difference is expected. The highest of these values belong to stages controlled by diffusion. Lin~et~al.\ find that oxygen diffusion sets the rate while the scale is compact, and that the rate is then independent of oxygen pressure at 873 to 973~K \cite{Lin2026}. Bartlett finds that the reaction at the buried boundary between carbide and oxide sets the rate between 723 and 853~K, and that the rate is also independent of oxygen pressure above about 5~mbar \cite{Bartlett1963}. In the simulations no such scale exists yet. Oxygen meets the carbide surface or an oxide a few atomic layers thick, and the rates depend on the oxygen available. Lin~et~al.\ find a similar situation once the scale has fully cracked. The reaction of the carbide then controls the rate, the rate depends on oxygen pressure, and the activation energy between 1073 and 1273~K falls to 14--38~kJ\,mol$^{-1}$ \cite{Lin2026}. The low values found here are thus in line with oxidation without a protective scale, although a similar value does not establish the same mechanism.

\begin{figure}[ht]
    \centering
    \includegraphics[width=\textwidth]{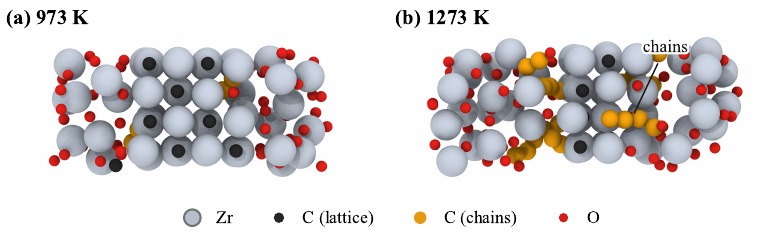}
    \caption{Illustrative final configurations at 140~ps. Zirconium is grey, lattice carbon black, chain carbon amber, and oxygen red. (a)~The 973~K cell with chain count nearest the four-cell median. (b)~The 1273~K cell with the largest chain reservoir. It holds 26 chain carbon atoms, against 12 to 16 in the other three cells, and is not representative of the average structure at this temperature. Carbon chains occupy the gap between the carbide and the displaced oxide.}
    \label{FigSimulationSnapshots}
\end{figure}

In our simulations, carbon accumulates while the post-40~ps fits suggest declining commitment rates. The fits do not establish whether retained carbon contributes to this slowdown. Barnier and Thevenot report substantial carbon retention in zirconium oxycarbides despite slightly lower oxidation resistance than zirconium carbide \cite{barnier1988comparative}. For a more developed layer, however, Gasparrini~\textit{et al.}\ attribute the oxidation rate at 1073~K to oxygen diffusion through the carbon--zirconia intermediate layer \cite{Gasparrini2018ZirconiumCarbide}. Both studies concern oxidation well beyond the stage of our cells, where the retained carbon is a few chains at each surface. Our simulations show how this carbon first collects, and its effect on the rate may only appear as the layer grows. The chains in our final structures occupy the gap between the carbide and displaced oxide (Fig.~\ref{FigSimulationSnapshots}), raising a further question about their effect on scale adhesion. Lin~\textit{et al.}\ associate $\mathrm{CO}_2$ formation with cracking and loss of scale protectiveness \cite{Lin2026}, whilst Bellucci~\textit{et al.}\ attribute gas evolution during cooling to cracks exposing retained carbon to oxygen \cite{Belluci2005}. Carbon gasification and scale damage may thus reinforce one another, with their sequence depending on the thermal and exposure history.

Our trajectories distinguish carbon bonded in chains from lattice carbon without carbon bonds. Experiments likewise need to identify the form of retained carbon as well as its amount. Bellucci~\textit{et al.}\ and Lin~\textit{et al.}\ combined gas analysis with spectroscopic and microscopic characterization \cite{Belluci2005, Lin2026}. Extending this approach with conversion-depth measurements in comparable regions would relate carbon accumulation to carbide consumption, while recording crack formation would help establish the sequence of carbon loss and scale damage. To test carbon's effect directly, simulations could compare interfaces that differ only in their carbon content. Tracking oxygen arrival and carbon liberation at the carbide would then test whether retained carbon limits oxygen access and slows conversion, while adhesion calculations would examine how it changes the energy needed to separate oxide from carbide.

Four independent trajectories at each temperature sample different reaction pathways under the gas-exchange protocol (Section~\ref{SecFluxControl}). This repetition is a central element of the small-cell approach, which replaces one large simulation by the statistical analysis of several small ones \cite{Hong2013SmallCell}. It compensates for the limited sampling of a single small cell and provides the sampling uncertainty. For melting points the approach is usually accurate to within 100~K in cells of more than 100 atoms \cite{Hong2013SmallCell}, although replication does not by itself establish convergence with cell size for oxidation. The small slab and short trajectories restrict the study to early growth; mature scale development and fracture require larger cells and longer observation times. Raising the cutoff or the k-point sampling changes the forces by 1 to 5\,\% (Section~\ref{SecConvergence}). The calculations use a stoichiometric (100) surface, so initial carbon vacancies and other facets are outside this study.

The high oxygen supply is what brings early oxidation within reach of 140~ps of \textit{ab initio} dynamics. It delivers in that time the oxygen that air at atmospheric pressure supplies in roughly 50~ns (Section~\ref{SecFluxControl}). The events observed agree with experiments at a far lower supply. Dosing experiments on $\mathrm{ZrC}(100)$ in vacuum likewise find dissociative uptake and a loss of surface carbon that is larger at 500~K than at room temperature \cite{Rodriguez2005}, and a disordered substoichiometric oxide \cite{Kato2000}. Experiments also retain less carbon as the oxygen pressure rises \cite{RamaRao1994, Bartlett1963}. The carbon retention found here is therefore not expected to be a product of the high supply, and may be lower than at atmospheric pressure. The rates of mass gain and carbon commitment, in contrast, are tied to the supply and to the prompt removal of gaseous products, and are specific to the simulated conditions. The calculations omit spin polarization, which has little effect on the solid away from adsorbing molecules but makes each arriving $\mathrm{O}_2$ molecule about 0.8~eV too high in energy (Section~\ref{SecSpin}). The rate of oxygen uptake, and the absolute rates of mass gain and carbon commitment that follow from it, may therefore be overestimated. The apparent activation energies and the partitioning of carbon between gas and chains also depend on the oxygen supply and may be affected as well, including the temperature at which carbon retention is lowest. The static checks give no sign that the oxide and the carbon chains themselves are described differently. Lower oxygen pressures require longer trajectories than DFT allows. Machine learning interatomic potentials could reach these times and show how the partitioning between gas and chains responds \cite{Li2024MLIP}.

In summary, we extended the small-cell approach from equilibrium phase coexistence to oxidation, by adding a gas exchange that supplies oxygen and removes the gaseous products. Sixteen trajectories of 140~ps follow the oxidation of $\mathrm{ZrC}$ well beyond the first surface reactions reached by earlier \textit{ab initio} studies. Under these conditions, the average rate of carbon conversion follows an Arrhenius trend, but its partitioning between gas and retained chains changes with temperature. Above 973~K, faster conversion increasingly produces retained carbon rather than additional gas. The approach is not specific to $\mathrm{ZrC}$, and such simulations could also be used to train machine learning interatomic potentials that reach lower oxygen pressures and longer times.

\section{Methods}\label{SecMethodology}

\subsection{DFT and AIMD setup}
\label{SecDFTAIMDSetup}

All calculations were performed using density functional theory (DFT) as implemented in the Vienna Ab initio Simulation Package (VASP) \cite{VASP}. The projector-augmented wave (PAW) method was used to describe electron-ion interactions, and the Perdew-Burke-Ernzerhof (PBE) generalized gradient approximation (GGA) \cite{PBE_Ref} was employed for the exchange-correlation functional. 

The Brillouin zone was sampled using a single special k-point at $(0.25, 0.25, 0.25)$ \cite{KPoint}. The plane-wave energy cutoff was set to $400$~eV. The electronic self-consistency loop was set to converge to a tolerance of $10^{-4}$~eV. Electronic occupancies were determined using Fermi-Dirac smearing, with the smearing width set to the thermal energy ($k_B T$) corresponding to each target simulation temperature (873--1273~K). Standard pseudopotentials were used for O ($2s^2 2p^4$) and C ($2s^2 2p^2$), while a semi-core potential was employed for Zr, treating 12 electrons as valence with a frozen $[\mathrm{Ar}] 3d^{10}$ core ($4s^2 4p^6 5s^2 4d^2$). Spin polarization was not included, and the effect of this choice is quantified in Section~\ref{SecSpin}.

To model the oxidation process, a simulation cell was constructed consisting of a solid $\mathrm{ZrC}$ slab and a gaseous $\mathrm{O}_2$ phase. A $3 \times 2 \times 2$ supercell of bulk $\mathrm{ZrC}$ (comprising 48~$\mathrm{Zr}$ and 48~$\mathrm{C}$ atoms) was cleaved along the (100) surface to create a $\sim 14.1$~\text{\AA} thick slab. A vacuum region twice the thickness of the slab was added to separate the periodic images, creating two exposed pristine surfaces. All atoms within the slab were allowed to move dynamically without any constrained bulk layers. This vacuum was subsequently filled with 10 $\mathrm{O}_2$ molecules (20~$\mathrm{O}$ atoms) aligned along the $x$-axis. Their initial locations were determined via a differential evolution optimization algorithm to maximize the minimum distance from any molecule to any other atom in the cell. 

All AIMD trajectories were propagated in the $NVT$ ensemble using a Nosé-Hoover thermostat to maintain the target temperatures of $873$~K, $973$~K, $1073$~K, and $1273$~K. While the continuous MD integration formally conserves the number of particles ($N$), our framework periodically handled the discrete addition of $\mathrm{O}_2$ and removal of carbon gases, effectively modeling an open, reactive system. The molecular dynamics time-step size was dynamically adjusted within a range of 0.8 to 1.6~fs by SLUSCHI \cite{Hong2013SmallCell}. For each temperature, 4 independent small cells were simulated for 140~ps. They differ in the initial positions of the $\mathrm{O}_2$ molecules and in the initial velocities.

\subsection{Effect of the non-spin-polarized treatment}
\label{SecSpin}

All trajectories were run without spin polarization, as in earlier AIMD of thermal oxidation \cite{Cvitkovich2023}. The carbide and the oxide are non-magnetic, earlier static calculations found no need for spin polarization for oxygen adsorbed on ZrC \cite{Rodriguez2005, Osei2016}, and spin polarization roughly doubles the cost of each electronic step. The triplet ground state of gas-phase $\mathrm{O}_2$ is therefore not represented.

\begin{figure}[ht]
    \centering
    \includegraphics[width=\textwidth]{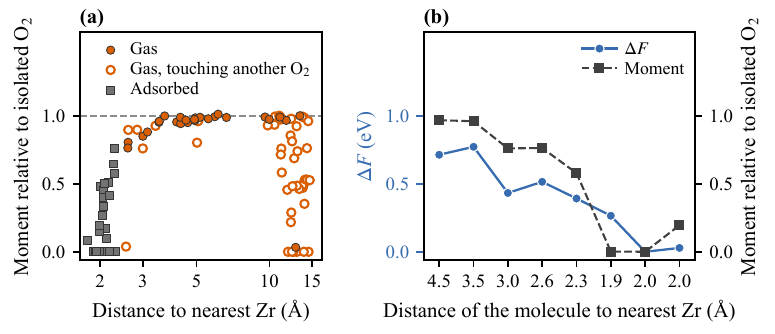}
    \caption{Effect of spin polarization on $\mathrm{O}_2$. (a)~Magnetic moment of each selected $\mathrm{O}_2$ molecule against its distance to the nearest Zr atom, for gas molecules (filled orange circles), gas molecules within 2.5~\text{\AA} of another $\mathrm{O}_2$ (open orange circles) and adsorbed molecules (grey squares). (b)~Difference in free energy between the non-spin-polarized and the spin-polarized calculation (blue circles, left axis) and moment of the molecule (black squares, right axis) along the adsorption of one $\mathrm{O}_2$ on the oxidized surface at 1273~K. The eight frames are in time order and labeled by the distance of the molecule to the nearest Zr atom. Moments are relative to the isolated molecule (dashed line in a).}
    \label{FigSpin}
\end{figure}

To measure the effect of this choice, we calculated 22 frames from the trajectories and an isolated $\mathrm{O}_2$ molecule with and without spin polarization, using the settings of the molecular dynamics with an electronic tolerance of $10^{-6}$~eV. The frames comprise four slabs at different stages of oxidation, two final structures with carbon chains, and eight frames along each of two adsorption events. Each spin-polarized calculation of a frame was started from two different initial spin states, and the result with the lower free energy was used. The energies of all structures are listed in Supplementary Section~S4. The largest magnetic moment on an atom of the solid is 0.10~$\mu_\mathrm{B}$, on a lattice carbon atom beside an adsorbing molecule. No Zr, chain carbon or oxide oxygen atom exceeds 0.04~$\mu_\mathrm{B}$. Where the $\mathrm{O}_2$ molecules carry almost no moment, the forces of the two treatments agree to within 0.005~eV~\text{\AA}$^{-1}$. Over all 22 frames they differ by 0.02 to 0.07~eV~\text{\AA}$^{-1}$ on the atoms of the solid, which is 1 to 3\,\% of the typical force of 2.2~eV~\text{\AA}$^{-1}$. The largest difference on an atom of the solid is 0.75~eV~\text{\AA}$^{-1}$, on a Zr atom beside an adsorbing molecule. A gas-phase $\mathrm{O}_2$ molecule carries 2~$\mu_\mathrm{B}$ and is 0.7 to 0.9~eV lower in free energy with spin polarization. This difference changes by less than 0.1~eV between $\mathrm{O}$--$\mathrm{O}$ distances of 1.15 and 1.70~\text{\AA}, so both treatments give the same bond length and stiffness. As a molecule approaches the surface, the moment and the energy difference decrease together (Fig.~\ref{FigSpin}a,b). Along an adsorption event on the oxidized surface, the difference is 0.7 to 0.8~eV in the gas, 0.4 to 0.5~eV at 2.6 to 3.0~\text{\AA} from the nearest Zr atom, and below 0.05~eV once the molecule is bound to the surface (Fig.~\ref{FigSpin}b). Each arriving molecule therefore releases up to about 0.8~eV more on adsorption than it would with spin polarization.

Without spin polarization, gas molecules also approach each other to within 1.7 to 2.3~\text{\AA}. For 20 such pairs, the force between the two molecules is attractive in 12 without spin polarization and repulsive in all 20 with it. About half of the gas molecules are in such a contact during the first 5~ps, while the ten initial molecules adsorb, and 12\% afterwards. The approximation thus acts mainly on the oxygen supply.

\subsection{Convergence with cutoff and k-point sampling}
\label{SecConvergence}

We tested the plane-wave cutoff and the k-point sampling on 14 of the frames of Section~\ref{SecSpin}, the four slabs, the two final structures with carbon chains, and the eight frames of the adsorption event on the oxidized surface. Each frame was recalculated with one setting changed, either a cutoff of 520~eV or a $\Gamma$-centered $1\times2\times2$ k-point mesh with one point along the surface normal. Four frames were also recalculated with a $1\times4\times4$ mesh. The results are given in Supplementary Section~S5.

Raising the cutoff from 400 to 520~eV changes the force on an atom by 0.02~eV~\text{\AA}$^{-1}$ (root mean square), which is 1\% of the typical force of 2.2~eV~\text{\AA}$^{-1}$ in these frames. The energy differences between the frames of the adsorption event, which span 5.4~eV, change by at most 0.03~eV.

The k-point sampling has a larger effect. Forces obtained with the single special k-point differ from those of the $1\times4\times4$ mesh by 0.08~eV~\text{\AA}$^{-1}$, or 4\% of the typical force, mainly on Zr and C atoms. The two meshes differ from each other by 0.04~eV~\text{\AA}$^{-1}$. The energy differences along the adsorption event change by at most 0.06~eV with the $1\times2\times2$ mesh. The difference between the frame with the molecule in the gas and the frame after adsorption is 4.01~eV with the special k-point and 4.03~eV with the $1\times4\times4$ mesh. For comparison, the thermal energy at the simulated temperatures is 0.075 to 0.110~eV. The single special k-point was used for its speed, as a calculation with the $1\times2\times2$ mesh takes about 4 times as long and one with the $1\times4\times4$ mesh about 11 times as long.

\subsection{Bond detection}
\label{SecBondDetection}

Atomic connectivity was determined using pair-specific distance cutoffs for gas species identification ($\mathrm{O}_2$, $\mathrm{CO}$, $\mathrm{CO}_2$) and coordination analysis. A pair was classified as bonded when its Cartesian separation, evaluated using the minimum-image convention for periodic boundary conditions \cite{LiquidsBook}, satisfied
\begin{equation}
    r_{ij} < r_{ij}^{\mathrm{cut}},
    \label{EqBondCutoff}
\end{equation}
where $r_{ij}^{\mathrm{cut}}$ is the cutoff for the species pair. The cutoffs were 2.31~\text{\AA} for $\mathrm{C}$--$\mathrm{C}$, 2.15~\text{\AA} for $\mathrm{C}$--$\mathrm{O}$, 3.33~\text{\AA} for $\mathrm{Zr}$--$\mathrm{C}$, 3.17~\text{\AA} for $\mathrm{Zr}$--$\mathrm{O}$, and 1.98~\text{\AA} for $\mathrm{O}$--$\mathrm{O}$. Each oxygen atom was restricted to a maximum of one $\mathrm{O}$--$\mathrm{O}$ bond, assigned by shortest pairwise distance.

\subsection{Dynamic gas exchange protocol}
\label{SecFluxControl}
 
Extending the equilibrium small-cell framework \cite{Hong2013SmallCell, Hong2015Highest} to oxidation requires three modifications. The system becomes open, with oxygen continuously supplied and gaseous products removed to prevent artificial back-pressure. The run tracks a sustained, irreversible transformation rather than an equilibrium endpoint, and the targets are mechanistic observables, oxide structure, product distribution, and rates, averaged over the trajectory ensemble.
 
To sustain the oxygen supply, the reactive gas-exchange protocol replenishes the $\mathrm{O}_2$ population using a threshold that tracks the available gas volume. We estimate the resulting impingement flux using the Hertz--Knudsen expression
\begin{equation}
J = \frac{N}{V_{\mathrm{gas}}} \sqrt{\frac{k_B T}{2 \pi m}}, \label{EqFlux}
\end{equation}
where $N$ is the number of free $\mathrm{O}_2$ molecules, those with no atom within 3.17~\text{\AA} of a Zr atom, $V_{\mathrm{gas}}$ the available gas volume, $m$ the $\mathrm{O}_2$ molecular mass, and $k_B$ the Boltzmann constant. The equivalent pressure form $J = P/\sqrt{2 \pi m k_B T}$ relates the estimated flux to the $\mathrm{O}_2$ partial pressures quoted alongside it. The gas volume is re-evaluated as the oxide grows, the instantaneous $\mathrm{O}_2$ count is smoothed to filter transient adsorption and desorption, and a molecule is added whenever the smoothed count falls below a threshold that tracks the shrinking gas volume, so the addition rate follows the receding gas phase rather than a fixed molecule count. The smoothing is an exponential average with a factor of 0.001 per MD step, which gives the count a memory of about 1~ps, and the threshold is half a molecule times the gas volume relative to its initial value. Newly added molecules are placed to maximize their distance from existing atoms. The threshold definitions and implementation parameters are given in Supplementary Section~S6. A related stepwise O$_2$ dosing scheme has been used to model the thermal oxidation of Si(100) \cite{Cvitkovich2023}. Carbide oxidation additionally requires the removal of volatile carbon oxides, which silicon oxidation does not produce.

\begin{figure}[ht]
    \centering
    \includegraphics[width=\textwidth]{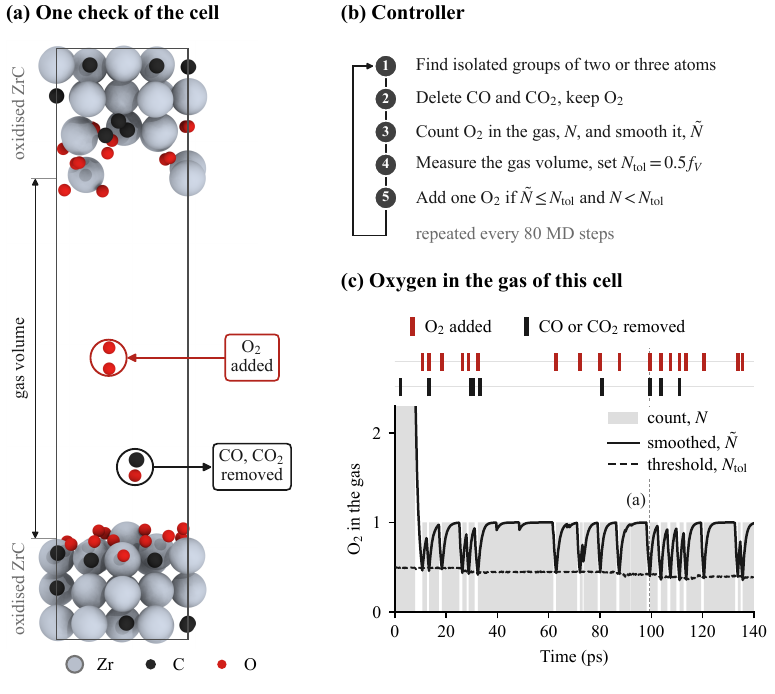}
    \caption{Gas exchange protocol. (a)~A cell at 873~K at the check at 99.4~ps. A detached $\mathrm{CO}$ molecule (black ring) is removed and an $\mathrm{O}_2$ molecule (red ring) is added at the position furthest from all atoms. The gas volume is measured between the outermost Zr atoms of the two surfaces. (b)~Steps of the controller. $f_V$ is the gas volume relative to its initial value. (c)~Additions of $\mathrm{O}_2$ and removals of $\mathrm{CO}$ or $\mathrm{CO}_2$ in the same cell, above the number of $\mathrm{O}_2$ molecules in the gas, $N$, its smoothed value $\tilde{N}$ and the threshold $N_\mathrm{tol}$. The dashed line marks the check shown in (a). The cell starts with 10 $\mathrm{O}_2$ molecules, and counts above 2 are not shown.}
    \label{FigGasExchange}
\end{figure}

Carbon-containing gas products ($\mathrm{CO}$ and $\mathrm{CO}_2$) are removed permanently (Fig.~\ref{FigGasExchange}). Every 80~MD steps, about 0.1~ps, the controller searches the current structure for isolated groups of two or three atoms. Groups other than $\mathrm{O}_2$ are deleted, and the oxygen count and the addition condition are then evaluated. Single unbound atoms are retained. In the trajectories, a product was removed at the first check at which it was detached by the criteria of Section~\ref{SecBondDetection} and its carbon atom was more than 3.0~\text{\AA} from every oxygen atom outside the molecule. For 37 of the 150 removed molecules this was one to three checks, at most 0.22~ps, after the molecule had first detached.

We examined every removal in the stored trajectories (Supplementary Section~S7). In the last frame before removal the molecules were at a median distance of 3.4~\text{\AA} from the nearest atom of the solid, 95\,\% were moving away from the surface, and their mean translational energy was 5.3~$k_BT$, against 2~$k_BT$ for thermal desorption. Removal does prevent some returns to the surface. Molecules that were detached at a check and stayed in the cell returned to the surface in 11 of 54 cases, and each of these carbon atoms left again as $\mathrm{CO}$ or $\mathrm{CO}_2$ within 1.1~ps. With all cutoffs shortened by 10\,\%, 24 further carbon atoms would have been removed at a check, and 22 of them were removed within 1.3~ps in the actual runs. The choice of criterion therefore shifts a removal by up to about 1~ps and changes the fate of at most 2 carbon atoms. Reactions of products that stay in the gas for longer are not represented.

The addition rule supplied 17 to 25 molecules per cell after the initial fill, one every 5.4 to 7.3~ps on average. The gas held one $\mathrm{O}_2$ molecule for 66 to 72\,\% of the time and none for 14 to 19\,\%. We did not repeat trajectories with other thresholds or smoothing constants, but their effect can be bounded from the rule itself. The threshold sets the number of molecules that the gas holds and therefore the pressure, to which the reported rates are tied. The smoothing constant sets the delay before an empty gas is refilled. The gas was empty for 14 to 19\,\% of the time, so a faster refill could raise the oxygen supply by about a fifth at most. This is comparable to the spread of the supply among the sixteen trajectories, which enters the bootstrap uncertainties.
 
The time-averaged impingement flux, averaged over the four trajectories at each temperature, ranged from $0.92$ to $1.04\times10^{29}$~m$^{-2}$~s$^{-1}$; individual trajectory averages ranged from $0.81$ to $1.18\times10^{29}$~m$^{-2}$~s$^{-1}$ (Supplementary Table~S11). For comparison, the Hertz--Knudsen flux of the oxygen in air at 1~bar is about $3\times10^{26}$~m$^{-2}$~s$^{-1}$ at these temperatures. The simulated supply is therefore at least about 300 times that of air and 3000 times that of a gas with 2\,\% $\mathrm{O}_2$. Air at 1~bar supplies the same amount of oxygen in roughly 50~ns. Depending on the temperature, an average of 43 to 62 oxygen atoms per cell have entered the oxide or left as $\mathrm{CO}$ and $\mathrm{CO}_2$ by 140~ps, or 2.7 to 3.9 per surface Zr atom. This is 81 to 90\,\% of the oxygen supplied. Rates are reported without flux normalization. As a sensitivity check, dividing the rates by the corresponding mean flux changed the apparent activation energies from 0.22 to 0.19~eV for net mass gain and from 0.16 to 0.13~eV for total carbon commitment, within their respective bootstrap uncertainties (Supplementary Section~S6).

\subsection{Bootstrap intervals for carbon partitioning}
\label{SecPartitioningMethods}

The confidence intervals and ordering probabilities of Section~\ref{sec:PhiC} come from a two-level bootstrap that mirrors the data hierarchy. Each carbon atom that is in a chain or has left as gas at 140~ps counts as one outcome. At the outer level the four trajectories per temperature are drawn with replacement, and at the inner level the carbon outcomes of each drawn trajectory are resampled with replacement from that trajectory's own outcomes, with the atom as the resampling unit. A drawn trajectory contributes as many outcomes as it originally had, and $\phi_\mathrm{C}$ is calculated from the outcomes of the four drawn trajectories together. Carbon atoms in one trajectory share the same oxide and history, so their outcomes are not independent, and whole trajectories are therefore redrawn at the outer level. Redrawing the trajectories alone gives narrower intervals at all four temperatures (Supplementary Table~S2b), so the wider intervals of the two-level scheme are reported. Intervals are the 2.5th and 97.5th percentiles of 10\,000 resampled $\phi_\mathrm{C}$ values, and every dominance or ordering statement quotes the fraction of resamples in which it holds, a resampled probability rather than a p-value. For the ordering statements, the four temperatures are resampled independently and compared in each resample. The scheme is seeded at 20260715, and the design rationale is given in Supplementary Section~S2.

\subsection{Competing-risks survival analysis of carbon commitment}
\label{SecSurvivalMethods}

The two properties that define the carbon states (Section~\ref{SecDesorptionKinetics}) give four combinations, which together with gas form five mutually exclusive states. In the starting structure each carbon atom has six zirconium neighbors within the $\mathrm{Zr}$--$\mathrm{C}$ cutoff, or five in the outermost carbon layers. A carbon atom keeps its zirconium coordination while at least five of these original neighbors remain within the cutoff, and it is bonded while another carbon atom lies within the $\mathrm{C}$--$\mathrm{C}$ cutoff (Section~\ref{SecBondDetection}). A change of either property is accepted once it has persisted for 0.2~ps (Supplementary Section~S9).

We estimated the rates in Section~\ref{SecDesorptionKinetics} using competing-risks survival analysis with cause-specific hazards. All 192 lattice carbons per temperature contribute their observed time before commitment, including those that remain uncommitted (Supplementary Section~S3). Chain commitment is timed at the first confirmed C--C bond, and gas commitment at the desorption time (Supplementary Section~S8). Carbon close to the coordination threshold can leave and rejoin the lattice repeatedly. To limit the effect of these repeated crossings, the liberation rate counts only liberations that are not reversed, in which the atom does not regain its zirconium coordination before the end of the simulation. Whether a liberation is reversed is known only from the rest of the trajectory.

Carbon uncommitted at 140~ps is censored at that time. The five atoms removed by a bookkeeping artifact (Supplementary Section~S8) are censored at their deletion times. Sensitivity of the rates to the event definition, the cutoffs, the coordination threshold and the persistence times is examined in Supplementary Section~S3.

Because gas and chain commitments drain the same pool of uncommitted carbon, the two cause-specific rates share one exposure and sum exactly to the total commitment rate. Each constant-rate estimate (Equation~\ref{eq:lambda}) carries an approximate model-based 95~\% interval $\left[\chi^2_{0.025}(2d)/2T_{\mathrm{exp}},\ \chi^2_{0.975}(2d)/2T_{\mathrm{exp}}\right]$, where $d$ counts the commitments and $T_{\mathrm{exp}}$ is the summed time at risk. These intervals assume a constant rate and independent carbon records; trajectory-level variation is assessed separately by bootstrap (Supplementary Section~S3). The committed fractions of Fig.~\ref{fig:Survival}c are cumulative incidence functions.

Time stationarity of each rate is tested using a Weibull hazard with shape $\rho$, fitted with lifelines 0.30.0 \cite{Davidson-Pilon2019}. To reduce the influence of the initial transient visible in Fig.~\ref{fig:Survival}c, we fit the interval from 40 to 140~ps. Only carbon still uncommitted and under observation at 40~ps is included, with 40~ps subtracted from subsequent event and censoring times. The constant-rate model is recovered at $\rho = 1$; shapes below or above one indicate declining or increasing rates per remaining uncommitted atom, respectively. The rate estimates and Arrhenius analysis in Fig.~\ref{fig:Survival}b use the full 140~ps window.

All estimates are deterministic given the event ledger, and the bootstrap probabilities of Supplementary Section~S3 use seed 20260715.

\subsection{Reconstruction of net mass gain}
\label{SecMacroObsMethods}

We calculated net mass change from oxygen retained in the scale and carbon lost as gas (Fig.~\ref{fig:MassFlows}). Each oxygen bound into the scale adds 15.999~amu and each carbon lost as gas removes 12.011~amu, values the text rounds to 16.0 and 12.0. An oxygen counts as bound into the scale when it belongs to a bonded component of the network of Section~\ref{SecBondDetection} that contains at least one zirconium, after intact O$_2$ molecules, identified by shortest-first unique O--O pairing below 1.98~\AA, are excluded. Because the trajectories carry different time stamps, each is interpolated onto a common 0.1~ps grid before the four cells are summed, and the summed series is smoothed with a 2~ps moving average to remove the flicker of transiently bonded atoms. Flow slopes are fitted on 20 to 140~ps, excluding the initial dose burst. The apparent activation energy is the slope of a straight-line fit of the logarithm of the four net rates against inverse temperature, without weights. The rates themselves involve no resampling. The uncertainty of the activation energy comes from redrawing the four trajectories at each temperature 10\,000 times and repeating the fit (Supplementary Section~S3).

\subsection{Density profiles and oxygen layers}
\label{SecProfilesMethods}

The density profiles are histograms of the atomic positions along the surface normal, measured from the Zr centre of mass, with a bin width of 0.25~\text{\AA} and frames 50~fs apart, averaged over the four trajectories of a temperature. Oxygen layers are the peaks of the oxygen profile outside the outermost carbide plane of each surface, and their positions are compared with the carbide planes continued outward (Supplementary Section~S1).

\backmatter

\section*{Data availability}
Product counts for the individual trajectories are provided in the supplementary information. The simulation inputs are available at \url{https://github.com/Philipwur/ZrC-Oxidation-Paper-Inputs}.

\section*{Code availability}
The oxidation framework is available at \url{https://github.com/Philipwur/SGUSCHI}. At the time of submission, the code supports VASP-based workflows for basic cubic structures. The input files of the simulations and the analysis scripts are available at \url{https://github.com/Philipwur/ZrC-Oxidation-Paper-Inputs}.

\section*{Acknowledgments}
This work was funded by QAIMS Lab support and the Delft AI Initiative, with contribution from the Faculty of Aerospace Engineering. We would like to thank the HPC team at the DelftBlue supercomputer for their support.

\section*{Author contributions}
PW, QJH, and YT designed the research and interpreted results; PW programmed the framework for the simulations and analyses with support from QJH; BC, MM, and SF contributed to methodological and writing review; PW wrote the manuscript with input from all authors.

\section*{Competing Interests}
The authors declare no competing interests.

\bibliography{sn-bibliography}

\end{document}